\documentclass[sigconf,authorversion,nonacm,screen]{acmart} 

\usepackage{algorithmic}
\usepackage{graphicx}
\usepackage{textcomp}
\usepackage{xcolor}
\usepackage{xspace}

\definecolor{light-gray}{gray}{0.8}
\definecolor{codegreen}{rgb}{0,0.6,0}
\definecolor{codegray}{rgb}{0.5,0.5,0.5}
\definecolor{codepurple}{rgb}{0.58,0,0.82}
\definecolor{backcolour}{rgb}{0.95,0.95,0.92}

\newcommand{\tool}[1]{\textsc{#1}\xspace}

\newcommand{\cpr}{\tool{CPR}}

\newcommand{\angelix}{\tool{Angelix}}
\newcommand{\prophet}{\tool{Prophet}}
\newcommand{\manybugs}{\tool{ManyBugs}}
\newcommand{\genprog}{\tool{GenProg}}
\newcommand{\fixtwofit}{\tool{Fix2Fit}}
\newcommand{\fonex}{\tool{f1x}}

\usepackage[colorinlistoftodos,prependcaption,textsize=tiny]{todonotes}

\usepackage{enumerate}
\usepackage[shortlabels]{enumitem}

\usepackage{tcolorbox}

\usepackage{multirow}
\usepackage[normalem]{ulem}
\useunder{\uline}{\ul}{}

\usepackage{listings}
\usepackage{etoolbox,environ}
\newtoggle{isArxivVersion}

\newcommand{\new}[1]{{#1}\xspace}

\newcommand{\newCheck}[1]{{#1}\xspace}

\copyrightyear{2022} 
\acmYear{2022} 
\setcopyright{acmlicensed}\acmConference[ICSE '22]{44th International Conference on Software Engineering}{May 21--29, 2022}{Pittsburgh, PA, USA}
\acmBooktitle{44th International Conference on Software Engineering (ICSE '22), May 21--29, 2022, Pittsburgh, PA, USA}
\acmPrice{15.00}
\acmDOI{10.1145/3510003.3510040}
\acmISBN{978-1-4503-9221-1/22/05}

\begin{document}

\title{Trust Enhancement Issues in Program~Repair}

\author{Yannic Noller}
\authornote{Joint first authors}
\affiliation{%
  \institution{National University of Singapore}
  \country{Singapore}}
\email{yannic.noller@acm.org}

\author{Ridwan Shariffdeen}
\authornotemark[1]
\affiliation{%
  \institution{National University of Singapore}
  \country{Singapore}}
\email{ridwan@comp.nus.edu.sg}

\author{Xiang Gao}
\authornote{Alternate email: gaoxiang9430@gmail.com}
\affiliation{%
  \institution{National University of Singapore}
  \country{Singapore}}
\email{gaoxiang@comp.nus.edu.sg}

\author{Abhik Roychoudhury}
\affiliation{%
  \institution{National University of Singapore}
  \country{Singapore}}
\email{abhik@comp.nus.edu.sg}

\begin{abstract}
Automated program repair is an emerging technology that seeks to automatically rectify bugs and vulnerabilities using learning, search, and semantic analysis. Trust in automatically generated patches is necessary for achieving greater adoption of program repair. Towards this goal, we survey more than 100 software practitioners to understand the artifacts and setups needed to enhance trust in automatically generated patches. Based on the feedback from the survey on developer preferences, we quantitatively evaluate existing test-suite based program repair tools. We find that they cannot produce high-quality patches within a top-10 ranking and an acceptable time period of 1 hour. The developer feedback from our qualitative study and the observations from our quantitative examination of existing repair tools point to actionable insights to drive program repair research. Specifically, we note that producing repairs within an acceptable time-bound is very much dependent on leveraging an abstract search space representation of a rich enough search space. Moreover, while additional developer inputs are valuable for generating or ranking patches, developers do not seem to be interested in a significant human-in-the-loop interaction.
\end{abstract}




\maketitle

\section{Introduction}
Automated program repair technologies \cite{LPR19} are getting increased attention. In recent times, program repair has found its way into the automated fixing of mobile apps in the SapFix project in Facebook \cite{sapfix}, automated repair bots as evidenced by the Repairnator project \cite{icse18repairnator}, and has found certain acceptability in companies such as Bloomberg \cite{Kirbas2021_APRBloomberg}. While all of these are promising, large-scale adoption of program repair where it is well integrated into our programming environments is considerably out of reach as of now. In this article, we reflect on the impediments towards the usage of program repair by developers. There can be many challenges towards the adoption of program repair like scalability, applicability, and developer acceptability. A lot of the research on program repair has focused on scalability to large programs and also to large search spaces \cite{prophet,angelix,fix2fit,sapfix}. Similarly, there have been various works on generating multi-line fixes \cite{angelix,extractfix}, or on transplanting patches from one version to another \cite{Shariffdeen2021_PatchTransplantation} --- to cover various use cases or scenarios of program repair.

Surprisingly, there is very little literature or systematic studies from either academia or industry on the developer trust in program repair. In particular, what changes do we need to bring into the program repair process so that it becomes viable to have conversations on its wide-scale adoption?
Part of the gulf in terms of lack of trust comes from a lack of specifications --- since the intended behavior of the program is not formally documented, it is hard to trust that the automatically generated patches meet this intended behavior. Overall, we seek to examine whether the developer's reluctance to use program repair may partially stem from not relying on automatically generated code. This can have profound implications because of recent developments on AI-based pair programming\footnote{Github Copilot \url{https://copilot.github.com/}},
which holds out promise for significant parts of coding in the future to be accomplished via automated code generation. 

In this article, we specifically study the issues involved in enhancing developer trust on automatically generated patches. Towards this goal, we first settle on the research questions related to developer trust in automatically generated patches. These questions are divided into two categories (a) expectations of developers from automatic repair technologies, and (b) understanding the possible shortfall of existing program repair technologies with respect to developer expectations. To understand the developer expectations from program repair, we outline the following research questions.

\begin{enumerate}[start=1,label={\bfseries RQ\arabic*},leftmargin=3em]
\new{\item To what extent are the developers interested to apply automated program repair (henceforth called APR), and how do they envision using it?}
\item Can software developers provide additional inputs that would cause higher trust in generated patches? If yes, what kind of inputs can they provide?
\item What evidence from APR will increase developer trust in the patches produced?
\end{enumerate}
For a comprehensive assessment of the research questions, we engage in both qualitative and quantitative studies. Our assessment of the questions primarily comes in three parts. To understand the developer expectations from program repair, we conduct a detailed survey (with 35 questions) among more than 100 professional software practitioners. Most of our survey respondents are developers, with a few coming from more senior roles such as architects. The survey results amount to both quantitative and qualitative inputs on the developer expectations since we curate and analyze respondents' comments on topics such as 
\new{the expected evidence for patch correctness provided by}
automated repair techniques. Based on the survey findings, we note that developers are largely open-minded in terms of trying out a small number of patches (no more than 10) from automated repair techniques, as long as these patches are produced within a reasonable time, say less than 1 hour. Furthermore, the developers are open to receiving specifications from the program repair method (amounting to evidence of patch correctness). They are also open-minded in terms of providing additional specifications to drive program repair. The most common specifications the developers are ready to give and receive are tests.

Based on the comments received from survey participants, we then conduct a quantitative comparison of certain well-known program repair tools on the widely used \manybugs benchmarks~\cite{manybugs}.  To understand the possible deficiency of existing program repair techniques with respect to outlined developer expectations as found from the survey, we formulate the following research questions.
\begin{enumerate}[start=4,label={\bfseries RQ\arabic*},leftmargin=3em]
\item Can existing APR techniques pinpoint high-quality patches in the top-ranking (e.g., among top-10) patches within a tolerable time limit \new{(e.g., 0.5/1/2 hours)}? 
\item What is the impact of additional inputs (say, fix locations and additional passing test cases) on the efficacy of APR?
\end{enumerate}
We note that many of the existing papers on program repair use liberal timeout periods to generate repairs, while in our experiments the timeout is strictly maintained at no more than one hour. We are also restricted to observing the first few patches, and we examine the impact of the fix localization by either providing and not providing the developer location. Based on a quantitative comparison of well-known repair tools \angelix~\cite{angelix}, \cpr~\cite{cpr}, \genprog~\cite{genprog}, \prophet~\cite{prophet} and \fixtwofit~\cite{fix2fit} --- we conclude that the search space representation has a significant role in deriving plausible/correct patches within an acceptable time period. In other words, an abstract representation of the search space (aided by constraints that are managed efficiently or aided by program equivalence relations) is at least as critical as a smart search algorithm to navigate the patch space. We discuss how the tools can be improved to meet developer expectations, either by achieving compilation-free repair or by navigating/suggesting abstract patches with the help of simple constraints (such as interval constraints).

Last but not the least, we note that program repair can be seen as automated code generation at a micro-scale. By studying the trust issues in automated repair, we can also obtain an initial understanding of trust enhancement in automatically generated code.

\section{Specifications in Program Repair}

The goal of APR is to correct buggy programs to satisfy given specifications. In this section, we review these specifications and discuss how they can impact patch quality.

\subsubsection*{Test Suites as Specification}
APR techniques such as \genprog~\cite{genprog} and \prophet~\cite{prophet} treat test suites as correctness specifications.
The test suite usually includes a set of passing tests and at least one failing test.
The repair goal is to correct the buggy program to pass all the given test suites.
Although test suites are widely available, they are usually incomplete specifications that specify part of the intended program behaviors. Hence, the automatically generated patch may overfit the tests, meaning that the patched program may still fail on program inputs outside the given tests. For instance, the following is a buggy implementation that copies $\mathtt{n}$ characters from source array $\mathtt{src}$ to destination array $\mathtt{dest}$, and returns the number of copied characters.
A buffer overflow happens at line 6 when the size of $\mathtt{src}$ or $\mathtt{dest}$ is less than $\mathtt{n}$.
By taking the following three tests (one of them can trigger this bug) as specification, a produced patch ($\mathtt{{++}index} {<} \mathtt{n}\mapsto$ $\mathtt{{++}index} {<} \mathtt{n}\ \&\&\ \mathtt{index} {<} 3$) can make the program pass the given tests. Obviously, the patched program is still buggy on test inputs outside the given tests.

\begin{lstlisting}[linewidth=\columnwidth,language=C]
int lenStrncpy(char[] src, char[] dest, int n){
    if(src == NULL || dest == NULL)
        return 0;
    int index = -1;
    while (++index < n)
        dest[index] = src[index]; // buffer overflow
    return index;
}
\end{lstlisting}

\begin{table}[h!]
\centering
\small
\begin{tabular}{lllrrr}
\toprule
\textbf{Type} & $\mathtt{\mathbf{src}}$ & $\mathtt{\mathbf{dest}}$ & $\mathtt{\mathbf{n}}$ & \textbf{Output} & \textbf{Expected Output}  \\
\midrule
Passing & SOF            & COM             & 3            & 3      & 3                \\
Passing & DHT            & APP0            & 3            & 3      & 3                \\
Failing & APP0           & DQT             & 4            & *crash & 3                \\
\bottomrule
\end{tabular}
\vspace{-3pt}
\end{table}

\subsubsection*{Constraints as Specification}
Instead of relying on tests, another line of APR research, e.g., \tool{ExtractFix}~\cite{extractfix} and CPR~\cite{cpr}, take constraints as correctness specifications. Constraints have the potential to represent a range of inputs or even the whole input space. Driven by constraints, the goal of APR is to patch the program to satisfy the constraints.
However, unlike the test suite, the constraints are not always available in practice; for this reason, techniques like \angelix~\cite{angelix} and SemFix~\cite{semfix} take tests as specifications but extract constraints from tests.
Certain existing APR techniques take as input coarse-grained constraints, such as assertions or crash-free constraints.
For instance, \tool{ExtractFix} relies on predefined templates to infer constraints that can completely fix vulnerabilities.
For the above example, according to the template for buffer overflow, the inferred constraint is
$\mathtt{index} {<} \mathtt{sizeof(src)} \&\& \mathtt{index} {<} \mathtt{sizeof(dest)}$.
Once the patched program satisfies this constraint, it is guaranteed that the buffer overflow is completely fixed.  Guarantees from such fixing of overflows/crashes do not amount to a guarantee of the full functional correctness of the fixed program.

\subsubsection*{Code Patterns as Specification}
Besides test suites and constraints, code patterns can also serve as specifications for repair systems.
Specifically, given a buggy program that violates a code pattern, the repair goal is to correct the program to satisfy the rules defined by the code pattern.
The code patterns can be manually defined~\cite{tan2016anti}, from static analyzers~\cite{van2018static}, automatically mined from large code repositories~\cite{bader2019getafix, bavishi2019phoenix}, etc.
Similar to the inferred constraints, code patterns cannot ensure functionality correctness.

\section{Survey Methodology}
Since constructing formal program specifications is notoriously difficult, the specifications used by APR tools cannot ensure patch correctness. Unreliable overfitting patches cause developers to lose trust in APR tools. 
This motivates us to enquire/survey developers on how APR can be enhanced to gain their trust.

We designed and conducted a survey with software practitioners, specifically to answer the first three research questions (RQ1-3).
In June 2021, we distributed a questionnaire to understand how developers envision the usage of automated program repair and what can be provided to increase trust in automatically generated patches.
Note that we followed our institutional guidelines and received approval from the Institutional Review Board (IRB) of our organization prior to administering the survey.

\subsubsection*{Survey Instrument} 
We asked in total 35 questions about how trustworthy APR can be deployed in practice. Our questions are structured into six categories:
\begin{enumerate}[start=1,label={\bfseries C\arabic*},leftmargin=2em]

\item \textit{Usage of APR} (RQ1): whether and how developers would engage with APR.

\item \textit{Availability of inputs/specifications} (RQ2): what kind of input artifacts developers can provide for APR techniques.

\item \textit{Impact on trust} (RQ2): how additional input artifacts would impact the trust in \new{auto-}generated patches.

\item \textit{Explanations} (RQ3): what kind of evidence/explanation developers expect for auto-generated patches.

\item \textit{Usage of APR side-products} (RQ3): what side-products of APR are useful for the developers, e.g., for manual bug-fixing.

\item \textit{Background}: the role and experience of the participants in the software development process.

\end{enumerate}
C1 will provide insights for RQ1, C2 and C3 for RQ2, and C4 and C5 for RQ3.
The questions are a combination of open-ended questions like \textit{"How would you like to engage with an APR tool?"} and close-ended questions like \textit{"Would it increase your trust in auto-generated patches if additional artifacts such as tests/assertions are used during patching?"} with Multiple Choice or a 5-point Likert scale.
The questionnaire itself was created and deployed with Microsoft Forms.
A complete list of our questions can be found in Table \ref{tab:questions} and in our replication package\iftoggle{isArxivVersion}{}{~\cite{artifact}}.

{
\footnotesize
\begin{table*}[]
\caption{Complete list of questions from the developer survey. In total 35 questions in 6 categories.}
\label{tab:questions}
\begin{tabular}{|p{1.8cm}|p{13cm}|p{1.95cm}|}
\hline
\textbf{Category} & \textbf{Question} & \textbf{Type} \\
\hline
\hline
 & Q1.1 Are you willing to review patches that are submitted by APR techniques? & 5-Point Likert Scale \\
 & Q1.2 How many auto-generated patches would you be willing to review before losing trust/interest in the technique? & Selection + Other… \\
C1 Usage of & Q1.3 How much time would you be giving to any APR technique to produce results? & Selection + Other… \\
APR & Q1.4 How much time do you spend on average to fix a bug? & Selection + Other… \\
 & Q1.5 Do you trust a patch that has been adopted from another location/application, where a similar patch was already accepted by other developers? & 5-Point Likert Scale \\
 & Q1.6 Would it increase your confidence in automatically generated patches if some kind of additional input (e.g., user-provided test cases) were considered? & 5-Point Likert Scale \\
 & Q1.7 Besides some additional input that is taken into account, what other mechanism do you see to increase the trust in auto-generated patches? & Open-Ended \\
 \hline
 & Q2.1 Can you provide additional test cases (i.e., inputs and expected outputs) relevant for the reported bug? & 5-Point Likert Scale \\
C2 Availability & Q2.2 Can you provide additional assertions as program instrumentation about the correct behavior? & 5-Point Likert Scale \\
of Inputs & Q2.3 Can you provide a specification for the correct behavior as logical constraint? & 5-Point Likert Scale \\
 & Q2.4 Would you be fine with classifying auto-generated input/output pairs as incorrect or correct behavior? & 5-Point Likert Scale \\
 & Q2.5 How many of such queries would you answer? & Selection + Other… \\
 & Q2.6 For how long would you be willing to answer such queries? & Selection + Other… \\
 & Q2.7 What other type of input (e.g., specification or artifact) can you provide that might help to generate patches? & Open-Ended \\
 & Q2.8 Please describe how you would like to engage with an APR tool. For example shortly describe the dialogue between you (as user of the APR tool) and the APR tool. Which input would you pass to the APR tool? What do you expect from the APR tool? & Open-Ended \\
 \hline
 & Q3.1 Would it increase your trust in auto-generated patches if additional artifacts such as tests/assertions are used during patching? & 5-Point Likert Scale \\
C3 Impact on trust & Q3.2 Which of the following additional artifacts will increase your trust? & Multiple Choice \\
 & Q3.3 What are other additional artifacts that will increase your trust? & Open-Ended \\
 \hline
 & Q4.1 Would it increase your trust when the APR technique shows you the code coverage achieved by the executed test cases that are used to construct the repair? & 5-Point Likert Scale \\
C4 Explanations for generated & Q4.2 Would it increase your trust when the APR technique presents the ratio of input space that has been successfully tested by the inputs used to drive the repair? & 5-Point Likert Scale \\
patches & Q4.3 What other type of evidence or explanation would you like to come with the patches, so that you can select an automatically generated patch candidate with confidence? & Open-Ended \\
 \hline
 & Q5.1 Which of the following information (i.e., potential side-products of APR) would be helpful to validate the patch? & Multiple Choice \\
C5 Usage of APR & Q5.2 What other information (i.e., potential side-products of APR) would be helpful to validate the patch? & Open-Ended \\
side-products & Q5.3 Which of the following information (i.e., potential side-products of APR) would help you to fix the problem yourself (without using generated patches)? & Multiple Choice \\
 & Q5.4 What other information (i.e., potential side-products of APR) would help you to fix the problem yourself (without using generated patches)? & Open-Ended \\
\hline
 & Q6.1 What is your (main) role in the software development process? & Selection + Other… \\
C6 Background & Q6.2 How long have you worked in software development? & Selection \\
 & Q6.3 How long have you worked in your current role? & Selection \\
 & Q6.4 How would you characterize the organization where you are employed for software development related activities? & Selection + Other… \\
 & Q6.5 What is your highest education degree? & Selection + Other… \\
 & Q6.6 What is your primary programming language? & Selection + Other… \\
 & Q6.7 What is your secondary programming language? & Selection + Other… \\
 & Q6.8 How familiar are you with Automated Program Repair? & 5-Point Likert Scale \\
 & Q6.9 Are you applying any Automated Program Repair technique at work? & Yes/No \\
 & Q6.10 Which Automated Program Repair technique are you applying at work? & Open-Ended \\
 \hline
\end{tabular}
\end{table*}
}

\subsubsection*{Participants} 
\begin{figure}
\centering
\includegraphics[width=\columnwidth]{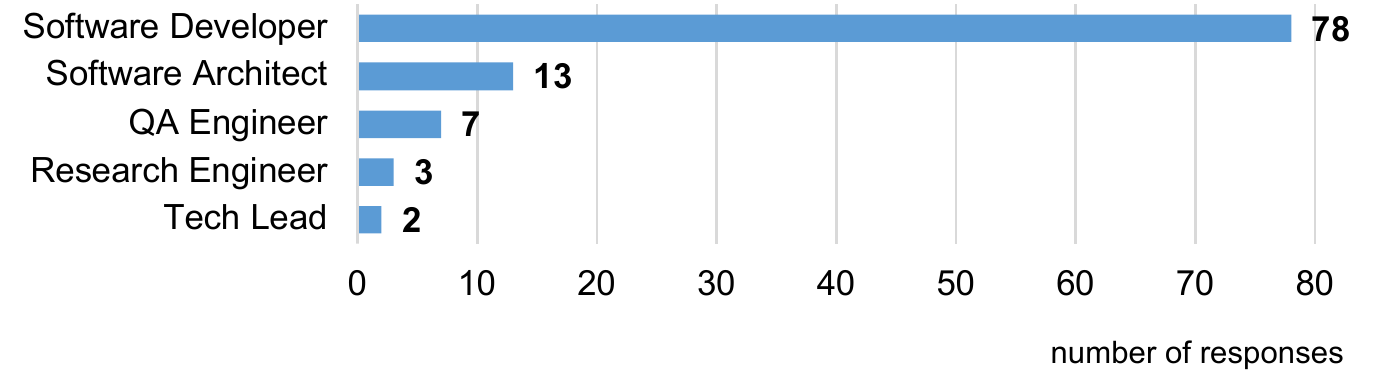}
\caption{Responses for Q6.1 \textit{What is your (main) role in the software development process?}}
\label{fig:roles}
\end{figure}

\begin{figure}
\centering
\includegraphics[width=\columnwidth]{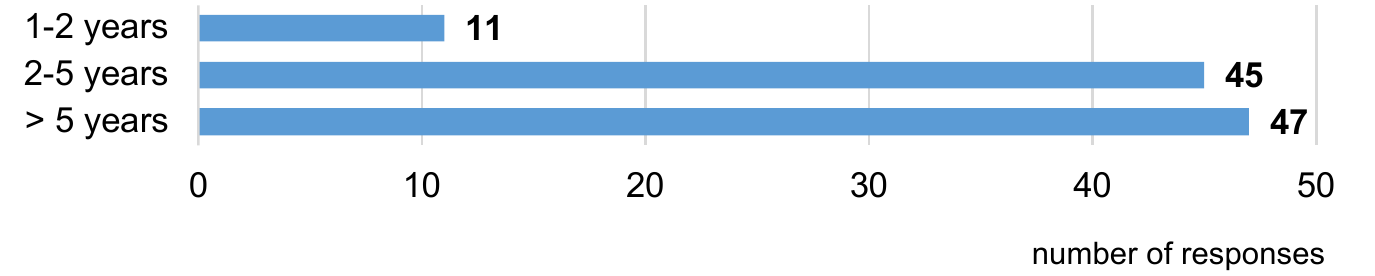}
\caption{Responses for Q6.2 \textit{How long have you worked in software development?}}
\label{fig:experience}
\end{figure}

We distributed the survey via two channels: (1) Amazon MTurk, and (2) personalized email invitations to contacts from global-wide companies.
As incentives, we offered each participant on MTurk 10 USD as compensation, while for each other participant, we donated 2 USD to a COVID-19 charity fund.
We received 134 responses from MTurk. To filter low-quality and non-genuine responses, we followed the known principles \cite{MTurkIssues} and used quality-control questions.
%
%
\new{In particular, we manually inspected all responses and filtered out answers that are irrelevant to the actual question: (1) we checked for suspicious answers, which overload keywords, e.g., many responses included a message on Annual Percentage Rate (APR) instead of automated program repair, and then
(2) we checked the consistency of the responses with quality-control questions, e.g., "Please describe briefly your role in software development" and "Name your primary activity in software development" at the beginning of the survey.
}
After this manual post-processing, we ended up with 34 \textit{valid} responses from MTurk.
From our company contacts, we received 81 responses, from which all have been genuine answers. 
From the total of 115 valid responses, we selected \textbf{103} \textit{relevant} responses, which excluded responses from participants who classified themselves as Project Manager, Product Owner, Data Scientist, or Researcher. Our goal was to include answers from software practitioners that have \new{daily,} hands-on experience in software development.
Figure \ref{fig:roles} and \ref{fig:experience} show the roles and experiences for the final subset of the 103 participants.

\subsubsection*{Analysis} 
For the questions with a 5-point Likert scale, we analyzed the distribution of negative (1 and 2), neutral (3), and positive (4 and 5) responses. For the Multiple Choice questions, we analyzed which choices were selected most, while the open-ended "Other" choices were analyzed and mapped to the existing choices or treated as new ones if necessary. For all open-ended questions, we performed a qualitative content analysis coding \cite{schreier2012qualitative} to summarize the themes and opinions.
The first iteration of the analysis and coding was done by one author, followed by the review of the other authors.
In the following sections, we will discuss the most mentioned responses, and indicate in the brackets behind the responses how often the topics are mentioned among the 103 participants.
\new{We use the chi-square goodness of fit test \cite{Pearson1900} ($\alpha=0.01$) to check that our results are significant and not a random observation. We also show the significance of the obtained trends/majorities with the Binomial Test \cite{BinomTest} ($\alpha=0.05$). We present the corresponding $P$ values.}
All data\new{, statistics}, and codes are included in our replication package\iftoggle{isArxivVersion}{}{~\cite{artifact}}.

\section{Survey Results}

\begin{figure*}
\centering
\includegraphics[width=1\textwidth]{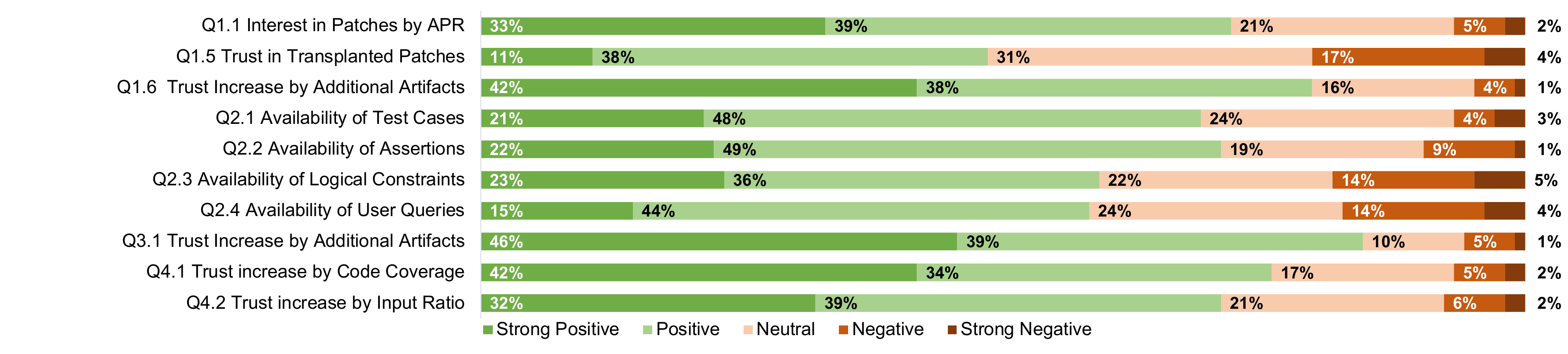}
\caption{Results for the questions with the 5-point Likert Scale (103 responses).}
\label{fig:results_likert}
\end{figure*}

\subsection{Developer engagement with APR (RQ1)}
In this section, we discuss the responses for the questions in category C1 and question Q2.8, which was explicitly exploring how the participants want to \textit{engage} with an APR tool.
First of all, a strong majority (72\% of the responses\new{, $P\!<\!.001$}) indicate that the participants are willing to review auto-generated patches (see Q1.1 in Figure \ref{fig:results_likert}). This finding generally confirms the efforts in the APR community to develop such techniques. Only 7\% of the participants are reluctant to apply APR techniques in their work.
\begin{figure}
\centering
\includegraphics[width=\columnwidth]{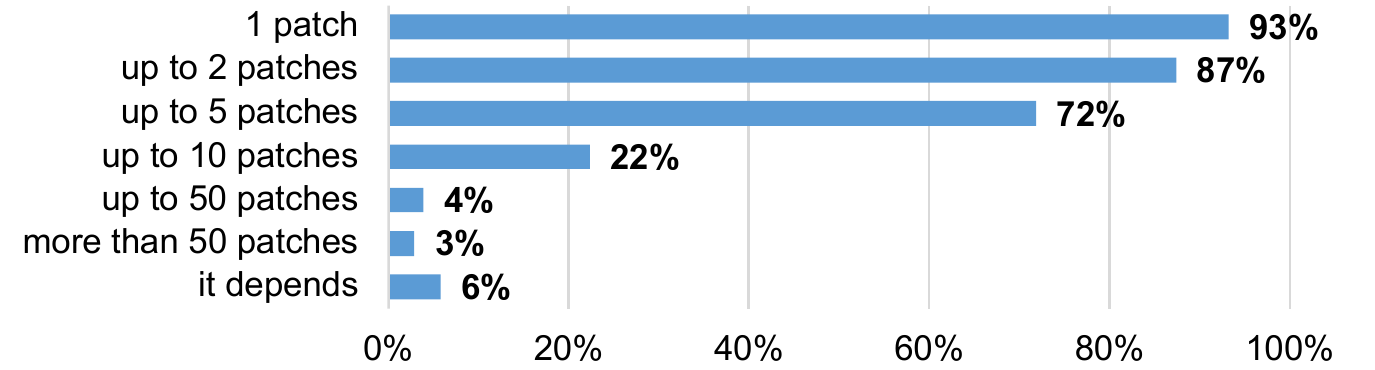}
\caption{Cumulative illustration of the responses for Q1.2 \textit{How many auto-generated patches would you be willing to review before losing trust/interest in the technique?}}
\label{fig:number_of_patches}
\end{figure}
As shown in Figure \ref{fig:number_of_patches}, we note that 72\% \new{($P\!<\!.001$)} of the participants want to review only up to 5 patches, while only 22\% would review up to 10 patches. Furthermore, 6\% mention that it would depend on the specific scenario.
%
At the same time, the participants expect relatively quick results: 63\% \new{($P\!=\!.003$)} would not wait longer than one hour, of which the majority (72\% of them, \new{$P\!<\!.001$}) prefer to not even wait longer than 30 minutes. The expected time certainly depends on the concrete deployment, e.g., repair can also be deployed along a nightly Continuous Integration (CI) pipeline, but our results indicate that direct support of manual bug fixing requires \textit{quick} fix suggestion or hints.
In fact, 82\% \new{($P\!<\!.001$)} of the participants state that they usually spend not more than 2 hours on average to fix a bug, and hence, the APR techniques need to be fast to provide a benefit for the developer.
To increase trust in the generated patches, 80\% \new{($P\!<\!.001$)} agree that \textit{additional artifacts} (e.g., test cases), which are provided as input for APR, are useful (see Q1.6 in Figure \ref{fig:results_likert}).
As a consistency check, we asked a similar question at a later point (see Q3.1 in Figure \ref{fig:results_likert}), and obtained that even \new{84\%} \new{($P\!<\!.001$)} agree that additional artifacts can increase trust.
The most mentioned \textit{other mechanisms} to increase trust are the \textit{extensive validation} of the patches with a test suite and static analysis tools (17/103), the actual \textit{manual investigation} of the patches (10/103), the \textit{reputation} of the APR tool itself (9/103), the \textit{explanation} of patches (8/103), and the provisioning of additionally \textit{generated tests} (7/103).

\begin{tcolorbox}[boxrule=1pt,left=1pt,right=1pt,top=1pt,bottom=1pt]
\textbf{RQ1 -- Acceptability of APR:}
Additional user-provided artifacts like test cases are \textit{helpful} to increase trust in automatically generated patches. However, our results indicate that \textit{full} developer trust requires a manual patch review. At the same time, \textit{test reports} of automated dynamic and static analysis, as well as \textit{explanations} of the patch, can facilitate the reviewing effort.
\end{tcolorbox}

The responses for the explicit question about developers' envisioned engagement with APR tools (Q2.8) can be categorized into four  areas: the extent of \textit{interaction}, the type of \textit{input}, the expected \textit{output}, and the expected \textit{integration} into the development workflow.

\subsubsection*{Interaction}
Most participants (71/103\new{, $P\!<\!.001$}) mention that they prefer a rather \textit{low} amount of interaction, i.e., after providing the initial input to the APR technique, there will be no further interaction. Only a few responses (6/103) mention the one-time option to provide more test cases or some sort of specification to narrow down the search space when APR runs into a timeout, or the generated fixes are not correct. Only 3 participants envision a high level of interaction, e.g., repeated querying of relevant test cases.

\subsubsection*{Input}
\new{Many} participants appear ready to provide \textit{failing test cases} (22/103) or \textit{relevant test cases} (20/103). Others mentioned that APR should take a \textit{bug report} as input (15/103), which can include the stack trace, details of the environment, and execution logs. Some also mentioned that they envision only the provision of the bare minimum, i.e., the program itself or the repository with the source code (11/103).

\subsubsection*{Output}
Besides the generated patches, the most mentioned helpful output from an APR tool is an \textit{explanation} of the fixed issue including its \textit{root cause} (9/103). This answer is followed by the requirement to present not only one patch but a \textit{list of potential patches} (8/103). Additionally, some participants mentioned that it would be helpful to produce a comprehensive \textit{test report} (6/103).

\subsubsection*{Integration}
The most mentioned integration mechanism is to involve APR smoothly in the \textit{DevOps pipeline} (17/103), e.g., whenever a failing test is detected by the CI pipeline, the APR would be triggered to generate appropriate fix suggestions. A developer would manually review the failed test(s) and the suggested patches. Along with the integration the participants mentioned that the primary goal of APR should be to \textit{save time} for the developers (8/103).

\begin{tcolorbox}[boxrule=1pt,left=1pt,right=1pt,top=1pt,bottom=1pt]
\textbf{RQ1 -- Interaction with APR:}
Developers envision a \textit{low} amount of interaction with APR, e.g., by only providing initial artifacts like test cases. APR should \textit{quickly} (within 30 min - 60 min) generate a \textit{small} number (between 5 and 10) of patches.
Moreover, APR needs to be \textit{integrated} into the existing DevOps pipelines to support the development workflow.
\end{tcolorbox}

\subsection{Availability/Impact of  Artifacts (RQ2)}
In this section, we look more closely in the categories C2 and C3 to investigate which additional artifacts can be provided by developers, and how these artifacts influence the trust in APR.
We first explore the availability of additional \textit{test cases} (69\% positive\new{, $P\!<\!.001$}), \textit{program assertions} (71\% positive\new{, $P\!<\!.001$}), and \textit{logical constraints} (59\% positive\new{, $P\!=\!.024$}) (see the results for Q2.1, Q2.2, and Q2.3 in Figure \ref{fig:results_likert}).
Furthermore, 58\% \new{($P\!=\!.038$)} of the participants are positive about answering queries to classify generated tests as failing or passing. This can be understood as participants want to have low interaction (i.e., asking questions to the tool), but if the tool is able to issue queries, they are ready to answer some of them (typically respondents prefer to answer no more than 10 queries\new{, $P\!=\!.001$}).
Based on the results for open-ended question Q2.7, the majority of the participants (70/103\new{, $P\!<\!.001$}) do not see any other additional artifacts (beyond tests/assertions/logical-constraints/user-queries) that they could provide to APR. The most mentioned responses by other participants are
different forms of \textit{requirements specification} (7/103), e.g., written in a domain-specific language,
\textit{execution logs} (6/103), 
documentation of \textit{interfaces} with data types and expected value ranges (5/103),
\textit{error stack traces} (4/103), 
relevant \textit{source code locations} (3/103), and reference solutions (3/103), e.g., existing solutions for similar problems. 

\begin{tcolorbox}[boxrule=1pt,left=1pt,right=1pt,top=1pt,bottom=1pt]
\textbf{RQ2 -- Artifact Availability:}
Software developers can provide additional artifacts like test cases, program assertions, logical constraints, execution logs, and relevant source code locations.
\end{tcolorbox}

\begin{figure}
\centering
\includegraphics[width=\columnwidth]{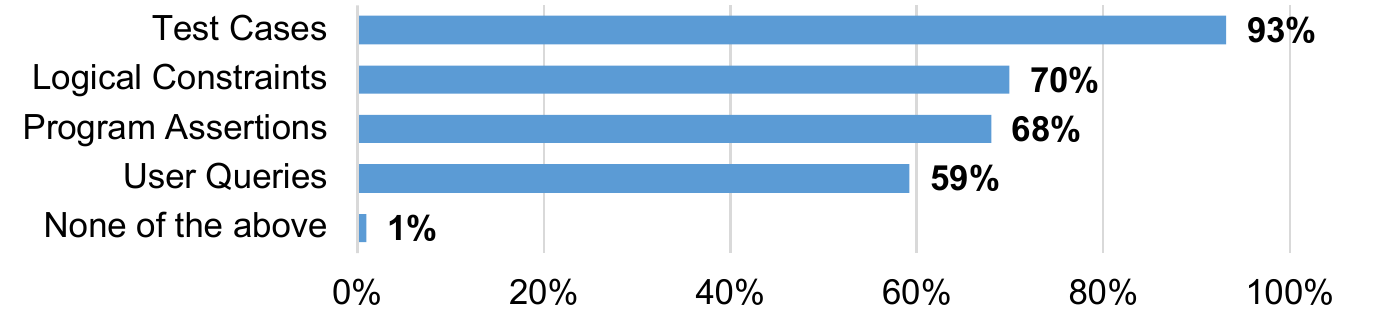}
\caption{Responses for Q3.2 \textit{Which of the following additional artifacts will increase your trust?}}
\label{fig:trust_artifacts}
\end{figure}

Regarding an \textit{increase} in trust in patches through the incorporation of additional artifacts driving repair, 93\% \new{($P\!<\!.001$)} of the participants agree that additional \textit{test cases} are helpful (see Figure~\ref{fig:trust_artifacts}). This is also interesting from the perspective of recent automated repair tools \cite{better,cpr} which perform automated test generation to achieve less overfitting patches. Logical constraints (70\%\new{, $P\!<\!.001$}) and program assertions (68\%\new{, $P\!<\!.001$}) perform worse in this respect. Although user queries allow more interaction with the APR technique, they would not necessarily increase trust more than the other artifacts.
\new{Only 59\% \new{($P\!=\!.024$)} agreed on their benefit.}
Most of the participants (88/103\new{, $P\!<\!.001$}) did not mention a trust gain by other artifacts. However, \new{some participants (3/103)} mentioned non-functional requirements like performance or security aspects, which is related to a concern that auto-generated patches may harm existing performance characteristics or introduce new security vulnerabilities.

\begin{tcolorbox}[boxrule=1pt,left=1pt,right=1pt,top=1pt,bottom=1pt]
\textbf{RQ2 -- Impact on Trust:}
Additional \textit{test cases} would have a great impact on the trustworthiness of APR. There exists the possibility of automatically generating tests to increase trust in \new{the auto-generated patches}.
\end{tcolorbox}

\subsection{Patch Explanation/Evidence (RQ3)}
In this section, we explore which patch evidence and APR side-products can support trust in APR (see categories C4 and C5).
We first proposed two possible pieces of evidence that could be presented along with the patches: the \textit{code coverage} achieved by the executed test cases that are used to construct the repair, and the \textit{ratio of input space} that has been successfully tested by the automated patch validation.
76\% \new{($P\!<\!.001$)} of the participants agree that code coverage would increase trust, and 71\% \new{($P\!<\!.001$)} agree with the input ratio (see Q4.1 and Q4.2 in Figure \ref{fig:results_likert}).
The majority of the participants (78/103\new{, $P\!<\!.001$}) do not mention other types of evidence that would help to select a patch with confidence. Nevertheless, the most mentioned response is a \textit{fix summary} (10/103), i.e., an explanation of what has been fixed including the root cause of the issue, how it has been fixed, and how it can prevent future issues. Other participants mention the \textit{success rate} in case of patch transplants (5/103), and a \textit{test report} summarizing the patch validation results (3/103). These responses match the observations for RQ1, where we asked how developers want to interact with trustworthy APR and what output they expect.

\begin{tcolorbox}[boxrule=1pt,left=1pt,right=1pt,top=1pt,bottom=1pt]
\textbf{RQ3 -- Patch Evidence:}
Software developers want to see \textit{evidence} for the patch's correctness to efficiently select patch candidates. 
Developers want to see information such as code coverage as well as the ratio of the covered input space.
\end{tcolorbox}

\begin{figure}
\centering
\includegraphics[width=0.85\columnwidth]{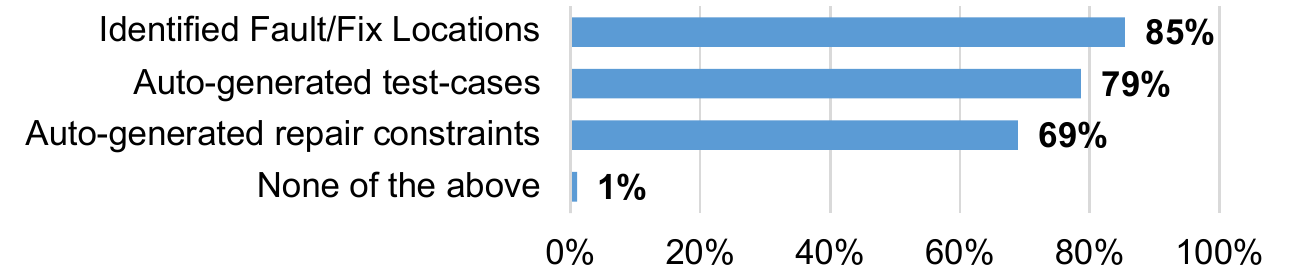}
\caption{Responses for Q5.1 \textit{Which of the following information (i.e., potential side-products of APR) would be helpful to validate the patch?}}
\label{fig:sideproducts_validation}
\end{figure}

A straightforward way to provide explanations and evidence is to provide outputs that are already created by APR as side-products. We listed some of them and asked the participants to select which of them would be helpful to validate the patches (see results in Figure \ref{fig:sideproducts_validation}). 85\% \new{($P\!<\!.001$)} agree that the identified \textit{fault} and \textit{fix locations} are helpful to validate the patch followed by the \textit{generated test cases} with 79\% \new{($P\!<\!.001$)} agreement.
In addition, a few participants emphasize the importance of a \textit{test report} (4/103), an explanation of the \textit{root cause} and the \textit{fix attempt} (4/103).

\begin{figure}
\centering
\includegraphics[width=0.85\columnwidth]{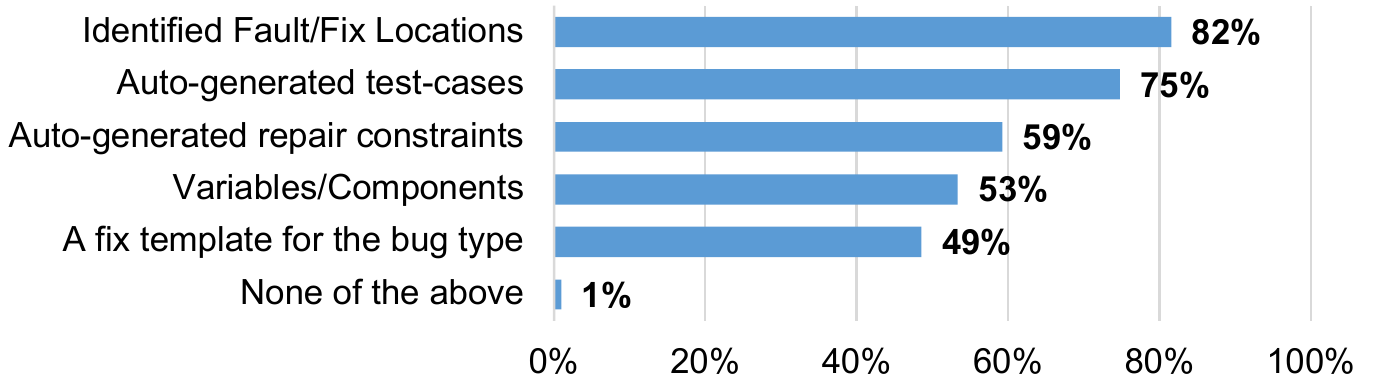}
\caption{Responses for Q5.3 \textit{Which of the following information (i.e., potential side-products of APR) would help you to fix the problem yourself (without using generated patches)?}}
\vspace*{-0.1in}
\label{fig:sideproducts_manualfix}
\end{figure}

Finally, we explore which side-products are most useful for developers, even when APR cannot identify the correct patch.
Figure \ref{fig:sideproducts_manualfix} shows that the identified fault and fix locations are of most interest (82\%\new{, $P\!<\!.001$}), followed by the generated test cases (75\%\new{, $P\!<\!.001$}).
Very few participants add that an issue summary (2/103) and the potential results of a data flow analysis (2/103) could be helpful too.

\begin{tcolorbox}[boxrule=1pt,left=1pt,right=1pt,top=1pt,bottom=1pt]
\textbf{RQ3 -- APR's Side-Products:}
Our results indicate that side-products of APR like the \textit{fault} and \textit{fix locations} and the \textit{generated test cases} can assist manual patch validation, and hence, enhance trust in APR.
\end{tcolorbox}

\section{Evaluation Methodology}
\label{sec:eval}
We now investigate to which extent existing APR techniques support the expectations and requirements collected with our survey. Not all aspects of our developer survey can be easily evaluated. For example, the evaluation of the amount of interaction, the integration into existing workflows, the output format for the efficient patch selection, and the patch explanations, require additional case studies and further user experiments.
In this evaluation, we focus on the quantitative evaluation of the relatively short patching time (30-60 min), the limited number of patches to manually investigate (5 to 10), handling of additional test cases and logical constraints, and the ability to generate a repair at a provided fix location. We explore whether state-of-the-art repair techniques can produce correct patches under configurations that match these expectations and requirements.
Specifically, we aim to provide answers to the research questions RQ4 and RQ5.

\subsubsection*{APR Representatives}
%
\new{In our evaluation, we selected tools to represent a wide spectrum of state-of-the-art APR techniques: search-based (\genprog~\cite{genprog}), semantic-based (\angelix~\cite{angelix}), the combination of search-based and learning-based (\prophet~\cite{prophet}), and the integration of testing inside repair to tackle overfitting (\fixtwofit~\cite{fix2fit}, \cpr~\cite{cpr}). We further selected tools that apply on C due to our evaluation subjects.
}
\genprog~\cite{genprog} is a search-based program repair tool that evolves the buggy program by mutating program statements. It is a well-known representative of the generate-and-validate repair techniques.
\angelix~\cite{angelix} is a semantic program repair technique that applies symbolic execution to extract constraints, which serve as a specification for subsequent program synthesis.
\prophet~\cite{prophet} combines search-based program repair with machine learning. It learns a code correctness model from open-source software repositories to prioritize and rank the generated patches.
\fixtwofit~\cite{fix2fit} combines search-based program repair with fuzzing. It uses grey-box fuzzing to generate additional test inputs to filter overfitting patches that crash the program. The test generation prioritizes tests that refine an equivalence class based patch space representation.
\cpr~\cite{cpr} uses semantic program repair and concolic test generation for refining abstract patches and for discarding overfitting patches. It takes a logical constraint as additional user input to reason about the generated tests inputs.

\subsubsection*{Subject Programs}
%
\new{We use the \manybugs~\cite{manybugs} benchmark, which is a well-established benchmark in APR, and all of the considered techniques/tools also use (some of) these subjects in their evaluation. Therefore, it is a benchmark for which it is known that the examined tools can identify patches. Our goal is to evaluate whether they can still identify patches with changed/limited environmental conditions (e.g., timeout, set of available test cases etc).
The benchmark set consists of 185 defects in 9 open-source projects. For each subject, \manybugs includes a test suite created by the original developers.
}
Note that all of the studied repair techniques require and/or can incorporate a test suite in their repair process. For our evaluation, we filter the 185 defects that have been fixed by the developer at a single fix location. We remove defects from "Valgrind" and "FBC" subjects due to the inability to reproduce the defects. Finally, we obtain 60 defects in 6 different open-source projects (see Table~\ref{table:subjects}).

\begin{table}[h]
\centering
\small
\caption{Experiment subjects and their details}
\label{table:subjects}
\begin{tabular}{llrcr}

\toprule
\textbf{Program} & \textbf{Description} & \textbf{LOC} & \textbf{Defects} & \textbf{Tests}  \\

\midrule
LibTIFF & Image processing library & ~77k & 7 & 78 \\
lighttpd & Web server & ~62k & 2 & 295\\
PHP & Interpreter & ~1046k & 43 & 8471\\
GMP & Math Library & ~145k & 1 & 146\\
Gzip & Data compression program & ~491k & 3 & 12\\
Python & Interpreter & ~407k & 4 & 355 \\
\bottomrule
\end{tabular}%
\end{table}

\subsubsection*{Experimental Configurations and Setup}
All tools are configured to run in full-exploration mode; which will continue to generate patches even after finding one plausible patch until the timeout or the completion of exploring the search space.
To study the impact of fix locations and test case variations (see RQ5), we evaluate each tool using different configurations (see Table~\ref{table:config}). In each configuration we provide the relevant source file to all techniques, however, with "developer fix location" we provide the exact source line number as well.
\new{Note that each setup uses a 1-hour timeout, which is chosen based on our survey responses: 63\% \new{($P\!=\!.003$)} of all participants would expect results within 1 hour.}

\begin{table}[h]
\centering
\small
\caption{Experiment configurations}
\label{table:config}
\begin{tabular}{llrr}
\toprule
\textbf{ID} & \textbf{Fix Location} & \textbf{Passing Tests} & \textbf{Timeout}  \\
\midrule
EC1 & tool fault localization & 100\% & 1hr \\
EC2 & developer fix location & 100\% & 1hr \\
EC3 & developer fix location & 0\% & 1hr \\
EC4 & developer fix location & 50\% & 1hr \\
\bottomrule
\end{tabular}%
\end{table}

\begin{table*}[ht]
\caption{Experimental results for the various configurations. Each cell shows the number of subjects, for which the technique was able to identify at least one \textit{Plausible}/\textit{Correct} patch with regard to the specific configuration. \newCheck{Please also see "Threats to Validity of Experimental Results" in Section \ref{sec:threats} to understand the context of these results fully.}}
\small
\begin{tabular}{|l|c|cccc|cccc|cccc|cccc|ccc|}
\hline
\multirow{2}{*}{\textbf{Subject}} &
\multirow{2}{*}{\textbf{Def.}} &
\multicolumn{4}{c|}{\textbf{\angelix}} &
\multicolumn{4}{c|}{\textbf{\prophet}}  &
\multicolumn{4}{c|}{\textbf{\genprog}} &
\multicolumn{4}{c|}{\textbf{\fixtwofit}} &
\multicolumn{3}{c|}{\textbf{\cpr}}
\\
 &  & EC1 & EC2 & EC3 & EC4 & EC1 & EC2 & EC3 & EC4 & EC1 & EC2 & EC3 & EC4 & EC1 & EC2 & EC3 & EC4 & EC2 & EC3 & EC4 \\
\hline
\hline
LibTIFF & 7 & 3/1 & 3/1 & 3/1 & 3/1 & 1/0 & 1/0 & 1/0 & 1/0 & 5/0 & 5/0 & 5/0 & 5/0 & 5/1 & 4/1 & 4/1 & 4/1 & 4/2 & 4/2 & 4/2  \\
lighttpd & 2 & - & - & - & - & 1/0 & 0/0 & 0/0 & 0/0 & 1/0 & 1/0 & 1/0 & 1/0 & 1/0 & 1/0 & 1/0 & 1/0 & - & - & -  \\
PHP & 43 & 0/0 & 0/0 & 0/0 & 0/0 & 0/0 & 0/0 & 2/1 & 3/1 & 0/0 & 0/0 & 10/1 & 0/0 & 8/1 & 4/2 & 7/2 & 5/1 & 5/4 & 5/4 & 5/4  \\
GMP & 1 & 0/0 & 0/0 & 0/0 & 0/0 & 0/0 & 0/0 & 0/0 & 0/0 & 0/0 & 0/0 & 0/0 & 0/0 & 0/0 & 0/0 & 0/0 & 0/0 & 1/1 & 1/1 & 1/1  \\
Gzip & 3 & 0/0 & 1/0 & 1/0 & 1/0 & 0/0 & 1/1 & 1/1 & 1/1 & 0/0 & 0/0 & 0/0 & 0/0 & 0/0 & 0/0 & 0/0 & 0/0 & 3/1 & 3/1 & 3/1  \\
Python & 4 & - & - & - & - & 0/0 & 1/1 & 1/1 & 1/1 & 0/0 & 0/0 & 0/0 & 0/0 & 0/0 & 0/0 & 0/0 & 0/0 & - & - & -  \\
\hline
\hline
Overall & 60 & 3/1 & 4/1 & 4/1 & 4/1 & 2/0 & 3/2 & 5/3 & 6/3 & 6/0 & 6/0 & 16/1 & 6/0 & 14/2 & 9/3 & 12/3 & 10/2 & 13/8 & 13/8 & 13/8  \\
\hline
\end{tabular}%
\label{table:results-all}
\end{table*}

\begin{table}[ht]
\caption{Experimental results for the average exploration ratio {$\mathbf{|P_{Expl}|}$} for EC1 and EC2.}
\small
\begin{tabular}{|l|rr|rr|rr|rr|}
\hline
\multirow{2}{*}{\textbf{Subject}} &
\multicolumn{2}{c|}{\textbf{\angelix}} &
\multicolumn{2}{c|}{\textbf{\prophet}}  &
\multicolumn{2}{c|}{\textbf{\genprog}} &
\multicolumn{2}{c|}{\textbf{\fixtwofit}} \\
 & EC1 & EC2 & EC1 & EC2 & EC1 & EC2 & EC1 & EC2 \\
\hline
\hline
LibTIFF & 86 & 100 & 24 & 93 & 1 & 27 & 100 & 100  \\
lighttpd & - & - & 20 & 100 & <1 & 51 & 100 & 100 \\
PHP & 96 & 100 & 22 & 96 & <1 & 91 & 63 & 80  \\
GMP & 100 & 100 & 41 & 100 & 5 & 100 & - & - \\
Gzip & 100 & 100 & 6 & 100 & 18 & 100 & 100 & 100 \\
Python & - & - & 14 & 100 & 1 & 100 & - & - \\
\hline
\hline
Overall & 95 & 100 & 21 & 98 & 4 & 78 & 91 & 95 \\
\hline
\end{tabular}%
\label{table:results-exploration}
\end{table}

\subsubsection*{Evaluation Metrics}
\new{In order to assess the techniques and support the answering of our research questions, we consider the following eight metrics, which are inspired by existing studies in APR \cite{Liu2021, Liu2020}:}
\textbf{M1} the search space \textit{size} of the repair tool,
\textbf{M2} the number of \textit{enumerated/explored} patches,
\textbf{M3} the explored \textit{ratio} with respect to the search space, 
\textbf{M4} the number of \textit{non-compilable} patches,
\textbf{M5} the number of \textit{non-plausible} patches, i.e., patches that have been explored but ruled out because existing or generated test cases are violated,
\textbf{M6} the number of \textit{plausible} patches,
\textbf{M7} the number of \textit{correct} patches,
and \textbf{M8} the highest \textit{rank} of a correct patch.
M1-M6 help to analyze the overall search space creation and navigation of each technique.
The definition of the search space size (M1) for the defect, as well as the definition of an enumerated/explored patch (M2), vary for each tool. We include all experiment protocols in our replication artifact, which describes how to collect these metrics for each tool.
M7-M9 assess the repair outcome, i.e., the identification of the \textit{correct} patch. We define a patch as \textit{correct} whenever it is \textit{semantically equivalent} to the developer patch \new{that is provided in} our benchmark.
To check for the correct patch, we manually investigated only the top-10 ranked patches because our survey concluded that developers would not explore beyond that. Note that not all techniques provide a patch ranking (e.g., \angelix, \genprog, and \fixtwofit).
\new{In these cases, we use the order of generation.}

\subsubsection*{Hardware}
All our experiments were conducted using Docker containers on top of AWS (Amazon Web Services) EC2 instances. We used the c5a.8xlarge instance type, which provides 32 vCPU processing power and 64GiB memory capacity.

\subsubsection*{Replication}
Our replication package contains all experiment logs and subjects, as well as protocols that define the methodology used to analyze the output of each repair tool\iftoggle{isArxivVersion}{}{~\cite{artifact}}. \new{In particular, we describe how to retrieve each evaluation metric for the specific repair techniques.}

\begin{tcolorbox}[boxrule=1pt,left=1pt,right=1pt,top=1pt,bottom=1pt]
\newCheck{
\textbf{Experimental Setup:}
Our experiments are meant to investigate specific aspects concerning the increase of \textit{program repair adoption} based on the results of our developer survey. We assume that the developer/user is not an APR expert, and hence, would use the \textit{default parameter} settings instead of fine-tuning or extending the tools.
Furthermore, our experiments use \textit{strict timeouts} and computation power restrictions. Other setups can lead to different and better results.
}
\end{tcolorbox}

\section{Evaluation Results}
Table~\ref{table:results-all} summarizes our evaluation results. For each APR technique we show its performance under the given experimental configuration (see Table \ref{table:config}). Each cell shows $|P_{Plaus}|$/$|P_{Corr}|$, where $|P_{Plaus}|$ is the number of defects for which the tool was able to generate at least one plausible patch (i.e., M6), and similarly $|P_{Corr}|$ is the number of defects for which the tool was able to generate a correct patch among the top-10 plausible patches.
For example, the LibTIFF project has 7 defects, for which \angelix was able to generate 3 plausible and 1 correct patch for the setup EC1 (i.e., 1-hour timeout, tool fault localization, and all available test cases).
Due to limitations in its symbolic execution engine KLEE~\cite{klee}, \angelix and \cpr do not support lighttpd and python, and the corresponding cells are marked with “-”. For \cpr, we are not able to produce results for EC1 because it does not have its own fault localization, and hence, requires the fix location as an input.
Additionally, Table~\ref{table:results-exploration} presents the average patch exploration/enumeration ratio $|P_{Expl}|$ of the techniques with respect to the patch space size, computed as a percentage of M2/M1 for each defect considered in each subject.

\subsection{APR within realistic boundaries (RQ4)}
\label{sec:rq4}
\new{The numbers in Table \ref{table:results-all} show that the overall repair success is limited.}
For example, \fixtwofit can generate plausible patches for 14 defects with EC1, while \cpr can generate correct patches for 8 defects given the correct fix location.
\new{Compared to previous studies, the number of plausible patches is lower in our experiments, mainly due to the 1-hour timeout}.
Prior research on program repair have experimented with 10-hour~\cite{f1x}, 12-hour~\cite{angelix, prophet} and 24-hour~\cite{fix2fit} timeouts, and determined whether a correct patch can be identified among \textit{all} generated plausible patches. The focus of these prior experiments was to evaluate the capability to generate a patch, whereas, in our work, we focus on the performance within a tolerable time limit set by developers.
\newCheck{
Not only the timeout but also a scenario-specific parameter fine-tuning can affect the results greatly. For example, when we modify the \textit{synthesis-level} parameter of \angelix (a parameter that modulates the back-end synthesis of the tool, and hence, can affect the search space), we can see additional patches being generated, such as 
for a defect in Libtiff (3edb9cd), in the EC3 configuration.
Our reported experiments only use the \textit{default parameters}. In future, for a full investigation of the repair tools' capabilities, it will therefore be necessary to conduct an exploration of the parameter choices in each repair tool, which has not been done in this paper.
}

\begin{tcolorbox}[boxrule=1pt,left=1pt,right=1pt,top=1pt,bottom=1pt]
\textbf{RQ4 -- Repair Success:}
\newCheck{Under our tight constraints (i.e., the 1-hour timeout and the top-10 ranking) and their default parameter setups, current state-of-the-art repair techniques cannot identify many plausible patches for the \manybugs benchmark.}
\end{tcolorbox}

\newCheck{Automated program repair tools are only beginning to gain adoption, and are still an emerging technology. We want to identify what it would take to increase the adoption of program repair.}
In general, the repair success of an APR technique is determined by (1) its search space, (2) the exploration of this search space, and (3) the ranking of the identified patches. In a nutshell, this means, if the correct patch is not in the search space, the technique cannot identify it. If the correct patch is in the search space, but APR does not identify it within a given timeout or other resource limitations, it cannot report it as a plausible patch. If it identifies the patch within the available resources but cannot pinpoint it in the (potentially huge) space of generated patches, the user/developer will not recognize it.
By means of these impediments for repair success in real-world scenarios, we examine the considered repair techniques. Our goal is to identify the concepts in APR that are necessary to achieve the developers' expectations, and hence, to improve the state-of-the-art approaches. 

\subsubsection*{Search Space}
Table~\ref{table:results-exploration} shows that \angelix explores almost its complete search space within the 1-hour timeout, while Table~\ref{table:results-all} shows that it can identify plausible patches for only one defect (with EC1). As described in~\cite{f1x}, the program transformations (to build/explore the search space) by \angelix only include the modification of existing side-effect-free integer expressions/conditions and the addition of if-guards. Therefore, we conclude that \angelix's search space is too limited to contain the correct patches.
The other techniques, on the other hand, consider larger search spaces. \prophet also considers the insertion of statements and the replacement of function calls. \genprog can insert/remove any available program statement. \fixtwofit uses the search space by \fonex~\cite{f1x}, which combines the search spaces of \angelix and \prophet to generate a larger search space.
\newCheck{\cpr uses the same program transformations as \angelix but is designed to easily incorporate additional user inputs like custom synthesis components to enrich its search space.}

\begin{tcolorbox}[boxrule=1pt,left=1pt,right=1pt,top=1pt,bottom=1pt]
\textbf{RQ4 -- Search Space:}
Successful repair techniques need to consider a wide range of program transformations and should be able to take \textit{user input} into account to enrich the search space.
\end{tcolorbox}

\subsubsection*{Search Space Exploration}
\prophet and \genprog show a \new{relatively} low exploration ratio with 21\% and 4\% respectively (see EC1 in Table~\ref{table:results-exploration}), which leads to a low number of plausible patches \new{in our experiments}.
Instead, \fixtwofit fully explores the patch search space for most of the considered defects (except for PHP), which leads to a high possibility of finding a plausible patch.
\cpr (not shown in the table) fully explores its search space in our experiments.
In contrast to \prophet and \genprog, \fixtwofit and \cpr perform \textit{grouping} and \textit{abstracting} of patches, to explore them efficiently.
\fixtwofit groups the patches by their behavior on test inputs and uses this equivalence relation to guide the generation of additional inputs.
\cpr represents the search space in terms of \textit{abstract} patches, which are patch templates, accompanied by constraints.
\cpr enumerates abstract patches instead of concrete patches, and hence, can reason about multiple patches at once to remove or refine patches. \prophet and \genprog, however, explore and evaluate all concrete patches, which causes a significant slowdown.
Reduction of the patch validation time is possible if we can validate patches without the need to re-compile the program for each concrete patch~\cite{Durieux2017_MetaProgramming, onthefly_patchvalidation, varfix}.

\begin{tcolorbox}[boxrule=1pt,left=1pt,right=1pt,top=1pt,bottom=1pt]
\textbf{RQ4 -- Patch Space Exploration:}
A large/rich search space requires an \textit{efficient} exploration strategy, which can be achieved by, e.g., using search space \textit{abstractions}.
\end{tcolorbox}

\subsubsection*{Patch Ranking}
Although \fixtwofit builds a rich search space and can efficiently explore it, it still cannot produce many correct patches \new{in our experiments}. One reason is that \fixtwofit can identify a correct patch but fails to pinpoint it in the top-10 patches because it only applies a rudimentary patch ranking, which uses the edit-distance between the original and patched program.
For instance, \fixtwofit generates the correct patch for the defect \textit{865f7b2} in the LibTiff subject but ranks it below position 10, and hence, it is not considered in our evaluation.
Furthermore, \fixtwofit's patch refinement and ranking is based on crash-avoidance, which is not suitable for a test-suite repair benchmark such as \manybugs that does not include many crashing defects.
\cpr improves on that by leveraging the user-provided logical constraint to reason about additionally generated inputs, while the patch behaviors on these inputs are collected and used to rank the patches. But still, overall, it cannot produce many correct patches within the top-10.
We also investigated how many of the correct patches are within the top-5 because 72\% \new{($P\!<\!.001$)} of our survey participants favored reviewing only up to 5 patches (see Figure \ref{fig:number_of_patches}). We observed that most identified correct patches within the top-10 are ranked very high so that there is not much difference if a top-5 threshold is applied.
Recent works \cite{Xiong2018_PatchCorrectness, varfix} propose the use of the test \textit{behavior} similarity between original/patched programs to rank plausible patches, which is a promising future direction.

\begin{tcolorbox}[boxrule=1pt,left=1pt,right=1pt,top=1pt,bottom=1pt]
\textbf{RQ4 -- Patch Ranking:}
After exploring the correct patch, an effective patch ranking is the last impediment for the developer.
\end{tcolorbox}

\subsection{Impact of additional inputs (RQ5)}

\subsubsection*{Providing Fix Location as User input}
In Table~\ref{table:results-all}, the column EC1 shows the results with the tool's fault localization technique, and column EC2 shows the results by repairing only at the developer-provided (correct) fix location.
Intuitively, one expects that equipped with the developer fix location, the results of each repair technique should improve.
However, the results by \angelix and \genprog do not change (except for one more plausible patch with \angelix).
From the previous discussion about the search space, we conclude that the program transformations by \angelix are the main limiting factor to the extent that even the provision of the correct fix location has no impact.
For \genprog we know from the EC3 configuration that there is at least one correct patch in the search space (see Table \ref{table:results-all}). Therefore, we conclude that \genprog suffers from its inefficient space exploration so that even the space reduction by setting the fix location has no impact.
\prophet instead can generate two additional correct patches in EC2, and hence, benefits from the precise fix location. The exploration ratio in Table \ref{table:results-exploration} shows that \prophet almost fully explores its search space in EC2, indicating a smaller search space.
\fixtwofit can generate one more correct patch as compared to EC1. Similar to \prophet, \fixtwofit benefits from the precise fix location and can explore more of its search space.
Note that \cpr is not included in the comparison between EC1 and EC2 because it does not apply for EC1. However, for EC2, it can generate the highest number of correct patches. Besides its efficient patch space abstraction, we attribute this to its ability to incorporate additional user inputs like the fix location and the user-provided logical constraint.

\begin{tcolorbox}[boxrule=1pt,left=1pt,right=1pt,top=1pt,bottom=1pt]
\textbf{RQ5 -- Fix Location:}
Our results show that the provision of the precise and correct fix location does not necessarily improve the outcome of the state-of-the-art APR techniques due to their limitations in search space construction and exploration. However, being amenable to such additional inputs can significantly improve the repair success, as shown by results from \cpr.
\end{tcolorbox}

\subsubsection*{Varying Passing Test Cases}
To examine the impact of the passing test cases, we consider the differences between the columns EC2, EC3, and EC4 in Table \ref{table:results-all}.
In general, more passing test cases can lead to high-quality patches because they represent information about the correct behavior. In line with this, we observe that more passing test cases lead to fewer plausible patches because the patch validation can remove more overfitting patches.
For \angelix however, we observe that there is no difference due to its limited search space.
\cpr is also not affected by the varying number of passing test cases. It uses the failing test cases to synthesize the search space and the passing test cases as seed inputs for its input generation. But since \cpr always fully explores the search space in our experiments, the variation of the initial seed inputs has no effect within the 1 hour.
Overall, we observe three different effects:
(a) For techniques with a limited search space (e.g., \angelix), passing test cases have very low or no effect.
(b) For techniques that suffer from inefficient space exploration strategies (e.g., \genprog and \prophet), having fewer passing test cases can speed up the repair process and lead to more plausible (possibly overfitting) patches.
(c) Otherwise (e.g., \fixtwofit), variations in the passing test cases can still influence the ranking.
Whether more tests are better depends on the APR strategy and its characteristics, as discussed in Section \ref{sec:rq4}. Therefore, we suggest that APR techniques incorporate an intelligent test selection or filtering mechanism, which is not yet studied extensively in the context of APR. Recently, \cite{Lou2021_TestSelection} suggested applying traditional regression test selection and prioritization to achieve better repair efficiency. Further developing and using such a mechanism represents a promising research direction.
Note that in the discussed experiments, the fix location was defined beforehand. However, if APR techniques use a test-based fault localization technique (like in EC1), the test cases have an additional effect on the search space and repair success.

\begin{tcolorbox}[boxrule=1pt,left=1pt,right=1pt,top=1pt,bottom=1pt]
\textbf{RQ5 -- Test Cases:}
Variation of passing test cases causes different effects depending on the characteristics of the APR techniques. Overall, one needs an intelligent test selection method.
\end{tcolorbox}

\section{Threats to Validity}
\label{sec:threats}

\subsubsection*{External Validity of Survey}
Although we reached out to different organizations in different countries, we cannot guarantee that our survey results can be generalized to all software developers. To mitigate this threat, we made all research artifacts publicly available\iftoggle{isArxivVersion}{}{~\cite{artifact}} so that other researchers and practitioners can replicate our study.
To reduce the risk of developers not participating or the volunteer bias, we designed the survey for a short completion time (15-20 min) and provided incentives like charity donations and (in the case of MTurk) monetary compensation.
\new{Another potential threat to validity is that only 15\% of all participants responded that they are \textit{familiar} with APR (see Q6.8/9/10). This is to be expected as APR is not (yet) heavily applied in the industry (with exceptions like Facebook and Bloomberg). To ensure that the participants have an idea of APR, we added a description and a link to an illustrative video at the beginning of our survey form. We note that we are exploring what it would take for developers to try out program repair, since developers may have general preconceived notions. By finding out what would make the developers comfortable to use APR, we can hope to increase adoption.}

\subsubsection*{Construct Validity of Survey}
In our survey, to encourage candid responses from participants, we did not collect any personally identifying information. Additionally, we applied control questions to filter non-genuine answers.
To mitigate the risk of wrong interpretation of the collected responses, we performed qualitative analysis coding, for which all codes have been checked and agreed by at least two authors.
Although we found general agreement across participants for many questions, we consider our results only as a first step towards exploring trustworthy APR. 

\subsubsection*{Internal Validity of Survey}
Our participants could have misunderstood our survey questions, as we could not clarify any particulars due to the nature of online surveys. To mitigate this threat, we performed a small pilot survey with five developers, in which we asked for feedback about the questions, the survey structure, and the completion time. Additionally, there is a general threat that participants could submit multiple responses because our survey was completely anonymous.

\newCheck{
\subsubsection*{Threats to Validity of Experimental Results}
In our empirical analysis, we do not cover all available APR tools, but instead, we cover the main APR concepts: search-based, semantics-based, and machine-learning-based techniques. With \manybugs~\cite{manybugs} we have chosen a benchmark that is a well-known collection of defects in open-source projects. Additionally, it includes many test cases, which are necessary to evaluate the aspects of test case provision. The metrics in our quantitative evaluation measure the patch generation progress, measuring repair efficiency/effectiveness via variations in configurations
(EC1-EC4).  To mitigate the threat of errors in our setup of experiments, we performed preliminary runs with a subset of the benchmark and manually investigated the results.

Our experimental results in Section \ref{sec:eval} explore the capability of the repair tools to produce patches within a 1-hour timeout. Different results may be observed if a different timeout is chosen. More importantly, it is possible to get  {\em significantly better results} from the repair tools by fine-tuning the parameters of the repair tools. For example, when we modify the \textit{synthesis-level} parameter of \angelix (a parameter that modulates the back-end synthesis of the tool, and hence, can affect the search space), we can see additional patches being generated, such as for a defect in Libtiff.
In our experiments, we did not fine-tune such parameters but instead used the default parameter settings, to simulate the experience of novice APR users. It is entirely possible that more expert APR users will be able to use the tools more effectively to get better results. The impact of parameter choices can also be rather nuanced e.g. Angelix is built on top of KLEE symbolic execution engine and KLEE has parameter settings of its own. 
Furthermore, we only share the results for the 1-hour timeout as it is closer to the time tolerance mentioned by our study participants.
}

\section{Related Work}
Our related work includes considerations of trust issues~\cite{Ryan2019_Trust, Alarcon2020_TrustRepair, Bertram2020_EyeTrackingTrust} and studies about the human aspects in automated program repair~\cite{Cambronero2019, Tao2014, liang2020interactive, Fry2012_PatchMaintainability, Kim2013_PatchAcceptability}, user studies about debugging \cite{Parnin2011}, and empirical studies about repair techniques~\cite{Liu2020, Kong2018, Motwani2020, Wang2019, wen2017empirical, Yang2020, Martinez2017, Liu2021}.
With regard to human aspects in automated program repair, our survey study contributes novel insights about the developers' expectations on their interaction with APR and which mechanisms help to increase trust.
With regard to empirical studies, our evaluation contributes a fresh perspective on existing APR techniques.

\subsubsection*{Trust Aspects in APR}
Trust issues in automated program repair emerge from the general trust issues in \textit{automation}. Lee and See~\cite{Lee2004_TrustAutomation} discuss that users tend to reject automation techniques whenever they do not trust them.
Therefore, for the successful deployment of automated program repair in practice, it will be essential to focus on its human aspects. 
With respect to this, our presented survey contributes to the knowledge base of how developers want to interact with repair techniques, and what makes them trustworthy.

Existing research on trust issues in APR focuses mainly on the effect of patch provenance, i.e., the source of the patch.
Ryan and Alarcon et al.~\cite{Ryan2019_Trust, Alarcon2020_TrustRepair} performed user studies, in which they asked developers to rate the trustworthiness of patches, while the researchers varied the source of the patches.
Their observations indicate that human-written patches receive a higher degree of trust than machine-generated patches. 
Bertram et al.~\cite{Bertram2020_EyeTrackingTrust} conducted an eye-tracking study to investigate the effect of patch provenance. They confirm a difference between human-written and machine-generated patches and observe that the participants prefer human-written patches in terms of readability and coding style.
Our study, on the other hand, explores the expectations and requirements of developers for trustworthy APR.
The work of Weimer et al.~\cite{Weimer2016_TrustedRepair} proposed strategies to assess repaired programs to increase human trust. Our study results confirm that an efficient patch assessment is crucial and desired by the developers.
We note that \cite{Weimer2016_TrustedRepair} focuses on how to assess APR, while we focus on how to enhance/improve APR in general, specifically in terms of its trust.

\subsubsection*{Human Aspects in APR}
Other human studies in the APR context focus on how developers interact with APR's output, i.e., the patches. 
Cambronero et al.~\cite{Cambronero2019} observed developers while fixing software issues.
They infer that developers would benefit from patch explanation and summaries to efficiently select suitable patches. They propose to explain the roles of variables and their relation to the original code, to list the characteristics of patches, and to summarize the effect of the patches on the program.
Tao et al.~\cite{Tao2014} explored how machine-generated patches can support the debugging process. They conclude that, compared to debugging knowing only the buggy location, high-quality patches can support the debugging effort, while low-quality patches can actually compromise it.
Liang et al.~\cite{liang2020interactive} concluded that even incorrect patches are helpful if they provide additional knowledge like fault locations.
Fry et al.~\cite{Fry2012_PatchMaintainability} explored the understandability and maintainability of machine-generated patches. While their participants label machine-generated patches as “slightly” less maintainable than human-written patches, they also observe that some augmentation of patches with synthesized documentation can reverse this trend.
Kim et al.~\cite{Kim2013_PatchAcceptability} proposed their template-based repair technique \tool{PAR} and evaluated the patch acceptability compared to \genprog.
All of these preliminary works explore the reactions on the output of APR. While our findings confirm previous hypotheses, e.g., that fault locations are helpful side-products of APR~\cite{liang2020interactive} or that an efficient patch selection is important~\cite{Weimer2016_TrustedRepair, liang2020interactive}, our work also considers the input to APR, the interaction with APR during patch generation, and how trust can be accomplished.

\subsubsection*{Debugging}
Parnin and Orso~\cite{Parnin2011} investigate the usefulness of debugging techniques in practice. They observe that many assumptions made by automated debugging techniques often do not hold in practice.
\new{Johnson et al.~\cite{Johnson2013} explore barriers for the wide adoption of static analysis tools and how well such tools fit into actual development workflows. They conduct interviews with developers and discuss their feedback to identify how those techniques can be improved.}
Although we focus on automated program repair, our research theme is related to \cite{Parnin2011} \new{and \cite{Johnson2013}}. We strive to understand how developers want to use automated program repair and whether current techniques support these aspects.

\subsubsection*{Empirical Evaluation of APR}
The living review article on automated program repair by Martin Monperrus \cite{repair-living-review} lists \new{(at the point of time we wrote this paper)} 43 empirical studies. Most of them are concerned about patch correctness to compare the success of APR techniques.
Other frequently explored aspects are repair efficiency \cite{Kong2018, Liu2020, Martinez2017, Liu2021}, the impact of fault locations \cite{Liu2020, wen2017empirical, Yang2020, Liu2021}, and the diversity of bugs \cite{Kong2018, Liu2021, Liu2021}.
Less frequently studied aspects are the impact of the test suite \cite{Kong2018, Motwani2020} and its provenance \cite{Motwani2020, Le2018}, specifically the problem of test-suite overfitting \cite{Le2018, Liu2021}, and how close the generated patches come to human-written patches \cite{Wang2019}.
Our empirical evaluation is not just another empirical assessment of APR technologies. It is specifically linked to the collected developer expectations from our survey. It limits the timeout to 1 hour, only explores the top-10 patches, and explores various configurations of passing tests as well as the impact of fix locations. Together with our survey results, our empirical/quantitative evaluation provides the building blocks to create trustworthy APR techniques, which will need to be validated via future user studies with practitioners.

\section{Discussion}
In this paper, we have investigated the issues involved in enhancing developer trust in automatically generated patches. Through a detailed study with more than 100 practitioners, we explore the expectations and tolerance levels of developers with respect to automated program repair tools. \newCheck{We then conduct a quantitative evaluation of existing repair tools to simulate the experience of novice APR users.} Our qualitative and quantitative studies indicate directions that need to be explored to gain developer trust in patches --- low interaction with repair tools, exchange of artifacts such as generated tests as inputs as well as output of repair tools, and paying attention to abstract search space representations over and above search algorithmic frameworks. \newCheck{ Each repair tool has many parameters and we only used the default parameter settings as would be expected from novice users  --- we did not explore the various parameter settings. To understand the full capability of the repair tools, in future it would be worthwhile to systematically explore a large number of parameter settings and try out the tools with various different timeouts.}

We note that increasingly there is a move towards automated code generation such as the recently proposed Github Copilot, but this raises the question of whether such automatically generated code can be trusted. Developing technologies to support mixed usage of manually written and auto-generated code, where program repair can improve the automatically generated code -- could be an enticing research challenge for the community.

\section*{Dataset from our work}
Our replication package with the survey and experiment artifacts is available on Zenodo \cite{artifact}.
\begin{center}
\href{https://doi.org/10.5281/zenodo.5908381}{\includegraphics[scale=0.25]{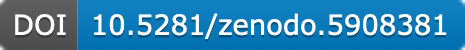}}
\end{center}

\new{
\section*{Acknowledgment}
This research is supported by the National Research Foundation, Prime Minister’s Office, Singapore under its Campus for Research Excellence and Technological Enterprise (CREATE) programme.
}

\bibliographystyle{ACM-Reference-Format}
\bibliography{references}

\end{document}